\documentclass[doublecol]{epl2} 

\title{The importance of scalar fields as extradimensional metric components in Kaluza-Klein models}
\shorttitle{The importance of scalar fields as extradimensional metric components in Kaluza-Klein models} 

\author{P. H. R. S. Moraes\inst{1} \and R. A. C. Correa\inst{1,2}}
\shortauthor{P. H. R. S. Moraes \etal}

\institute{                    
  \inst{1} ITA - Instituto Tecnol\'ogico de Aeron\'autica - Departamento de F\'isica - S\~ao Jos\'e dos Campos, S\~ao Paulo, Brazil\\
  \inst{2} SISSA - Scuola Internazionale Superiore di Studi Avanzati - Via Bonomea, 265, I-34136 Trieste, Italy
}
\pacs{11.10.Kk}{First pacs description}
\pacs{04.50.Cd}{Second pacs description}
\pacs{04.20.Cv}{Third pacs description}

\abstract{
Extradimensional models are achieving their highest popularity nowadays, among other reasons, because they can plausible explain some standard cosmology issues, such as the cosmological constant and hierarchy problems. In extradimensional models, we can infer that the four-dimensional matter rises as a geometric manifestation of the extra coordinate. In this way, although we still cannot see the extra dimension, we can relate it to physical quantities that are able to exert such a mechanism of matter induction in the observable universe. In this work we propose that scalar fields are those physical quantities. The models here presented are purely geometrical in the sense that no matter lagrangian is assumed and even the scalar fields are contained in the extradimensional metric. The results are capable of describing different observable cosmic features and yield an alternative to ultimately understand the extra dimension and the mechanism in which it is responsible for the creation of matter in the observable universe.}

\begin{document}

\maketitle

\section{Introduction}\label{sec:int}

Extra dimensions have been an useful artefact in Physics for a long time. Five years after the appearance of Einstein's General Relativity (GR), T. Kaluza has firstly expanded it to encompass an extra dimension \cite{kaluza/1921}. Kaluza could describe both four-dimensional (4D) GR and Maxwell's electromagnetism equations from an appropriate choice for the five-dimensional (5D) metric, in which the extra terms depend on the electromagnetic potential $A_a$ as \cite{overduin/1997}

\begin{equation}\label{i1}
g_{AB}=\left({\begin{array}{cc}
   g_{ab}+A_a A_b\phi^{2} & A_a \phi^{2} \\ A_b\phi^{2} & \phi^{2} \end{array}} \right),
\end{equation}
with $A,B$ running from $0$ to $4$, $a,b$ running from $0$ to $3$ and $\phi$ being a scalar field. Such an achievement can be made by substituting (\ref{i1}) in the 5D version of GR field equations for vacuum, i.e., 

\begin{equation}\label{i2}
G_{AB}=0.
\end{equation}
In this way, for Kaluza, matter in 4D space is a purely geometric manifestation of a 5D empty space. This confirmation agrees with a goal once espoused by Einstein, that {\it matter comes from geometry}, and shall be revisited and clarified.

Since Kaluza's work, the extradimensional concept has been applied to gravitational waves \cite{ponce_de_leon/2003,mm/2014}, gravitational lensing \cite{sadeghi/2013} and cosmology \cite{mc/2016}, among many other areas. 

In the present article, the Kaluza's concept of extra dimension will be taken into account. On this regard, it should be reminded that in the year of 1926, O. Klein have significantly contributed to Kaluza's idea \cite{klein/1926}. In order to account for the fact that apparently we live in a 4D world, Kaluza has proposed the so-called ``cylindrical condition'', which consists in the annulment of all derivatives with respect to the extra coordinate. Klein has made the cylindrical condition less artificial, by arguing that the annulment of the derivatives is a consequence of the extra dimension compactification.

An elegant, simple and useful form of physically interpreting Kaluza-Klein (KK) model came from a series of works from P.S. Wesson and collaborators \cite{wesson/1992,wesson/1992b,ponce_de_leon/1993,wesson/1996,liu/1998,fukui/1999,wesson/2000,wesson/2015,moraes/2016}. Wesson's idea, which gave rise to the so-called ``Induced Matter Model'' (IMM) (sometimes referred to as ``Space-Time-Matter Model''), was to collect all the terms coming from the extra dimension in Eq.(\ref{i2}) and make them pass to its {\it rhs} as an (induced) energy-momentum tensor.

The IMM has been widely applied to cosmology, yielding interesting and testable results \cite{overduin/2007,ponce_de_leon/2010,ponce_de_leon/2010b,romero/2013,moraes/2014,moraes/2015}.

In what concerns 5D models, a novel form of understanding the extra dimension is being established \cite{overduin/2017}. It is stated that we can infer from the extra dimension consequences that it has to do with the origin of matter and particle masses - though to fully understand these will require a better knowledge of the scalar field associated with the extra coordinate. The extra dimension should be accepted, then, in this context, as an useful concept.

From the above paragraph perspective, it is plausible to see the extra dimension as directly related with a scalar field, with this scalar field being responsible for inducing matter in the 4D universe. This is the main goal of the present article, in which the models will be constructed from a purely geometric approach, that is, the scalar fields will be considered as a part of the 5D metric in empty space.

Recently, the Higgs field existence was confirmed in laboratory \cite{aad/2012,chatrchyan/2012}, which has optimized the possibility of existence of other scalar fields, such as quintessence for example, which are named scalar fields responsible for the Universe dynamics \cite{caldwell/1998,ms/2014,khurshudyan/2015,chaubey/2016}. Scalar field models are also approached in the study of false vacuum transitions \cite{coleman/1977,callan/1977}, which, for instance, concern to statistical mechanics and also cosmology, the latter because the transition from false to true vacua can be interpreted as transitions between different stages of the Universe dynamics \cite{cmr/2015}.

\textcolor{black}{Another contexts where scalar fields play a key role can be found during the generation of coherent structures after cosmic inflation \cite{gleiser/2011}, in Lorentz and CPT breaking systems \cite{brito/2012, rafael/2015}, in regular asymptotically AdS reflecting star backgrounds \cite{yan/2018}, in hydrodynamic fluctuations \cite{jackson/2018} and in the production of gravitational waves during preheating after inflation \cite{stefan/2017}.}

In the next section we will start relating directly a scalar field with the extra dimension by applying the IMM for two different cases of (\ref{i1}). My proposal is to construct models that relate different scalar fields with the extra dimension and analyse the referred consequences.

\section{Induced Matter Model applications}\label{sec:immhf}

In this section we will consider two different cases of KK metric (\ref{i1}) and apply the IMM to them. In both cases, the electromagnetic potential $A_a$ will be neglected.

\subsection{Case I: $a=a(t)$ and $\phi=\phi(t)$}\label{ss:pt}

We are going to start by considering the simplest case for the scalar field of Eq.(\ref{i1}), that is, $\phi=\phi(t)$. The line element can then be written as

\begin{equation}\label{imm1}
dS^{2}=dt^{2}-a(t)^{2}(dx^{2}+dy^{2}+dz^{2})-\phi(t)^{2}dl^{2},
\end{equation}
in which $g_{ab}$ was taken as the Friedmann-Robertson-Walker metric with null curvature (in accord with observations \cite{hinshaw/2013}), $a(t)$ is the scale factor, which dictates how distances grow in an expanding universe, $l$ is the extra coordinate and throughout this work, it will be assumed units such that $8\pi G=c=1$.

Let us apply the IMM to Eq.(\ref{imm1}). This consists in evaluate (\ref{i2}), collect all the extradimensional terms and relate them with the matter content of the observable universe; that is, matter-energy density $\rho$ and pressure $p$.

The Einstein tensor for (\ref{imm1}) reads

\begin{equation}\label{imm2}
G_0^{0}=3\left(\frac{\dot{a}}{a}\right)^{2}+3\frac{\dot{a}}{a}\frac{\dot{\phi}}{\phi},
\end{equation}
\begin{equation}\label{imm3}
G_1^{1}=2\frac{\ddot{a}}{a}+\left(\frac{\dot{a}}{a}\right)^{2}+\frac{\ddot{\phi}}{\phi}+2\frac{\dot{a}}{a}\frac{\dot{\phi}}{\phi},
\end{equation}
\begin{equation}\label{imm4}
G_2^{2}=G_3^{3}=G_1^{1},
\end{equation}
\begin{equation}\label{imm5}
G_4^{4}=3\left[\frac{\ddot{a}}{a}+\left(\frac{\dot{a}}{a}\right)^{2}\right],
\end{equation}
in which dots represent time derivatives. We consider, as in braneworld models, that one cannot find matter in the extra dimension, that is, the energy-momentum tensor of a perfect fluid is written as $T_{B}^{A}=diag(\rho,-p,-p,-p,0)$. 

Eqs.(\ref{imm2}), (\ref{imm3}) and (\ref{imm5}) yield, respectively

\begin{equation}\label{imm5.1}
\frac{\dot{a}}{a}=-\frac{\dot{\phi}}{\phi},
\end{equation}
\begin{equation}\label{imm5.2}
2\frac{\ddot{a}}{a}-\left(\frac{\dot{a}}{a}\right)^{2}=-\frac{\ddot{\phi}}{\phi},
\end{equation}
with $\rho=-3(\dot{a}/a)(\dot{\phi}/\phi)$, $p=2(\dot{a}/a)(\dot{\phi}/\phi)+\ddot{\phi}/\phi$ and the trivial differential equation

\begin{equation}\label{imm6}
\frac{\ddot{a}}{a}+\left(\frac{\dot{a}}{a}\right)^{2}=0,
\end{equation}
whose solution reads 

\begin{equation}\label{imm7}
a(t)=\alpha_1\sqrt{2t-\beta_1},
\end{equation}
with $\alpha_1$ and $\beta_1$ being constants.

By substituting (\ref{imm7}) in (\ref{imm5.1})-(\ref{imm5.2}) and rearranging them, yields

\begin{equation}\label{imm8}
\frac{\ddot{\phi}}{\dot{\phi}}=\frac{3}{\beta_1-2t}.
\end{equation}

The solution for (\ref{imm8}) is

\begin{equation}\label{imm9}
\phi(t)=\gamma_1-\frac{\delta_1}{\sqrt{2t-\beta_1}},
\end{equation}
with $\gamma_1$ and $\delta_1$ constants.

Fig.\ref{fig1} below shows the time evolution of $\phi(t)$ for different values of the constants involved. 

\begin{figure}[ht!]
\vspace{0.3cm}
\centering
\includegraphics[height=5cm,angle=00]{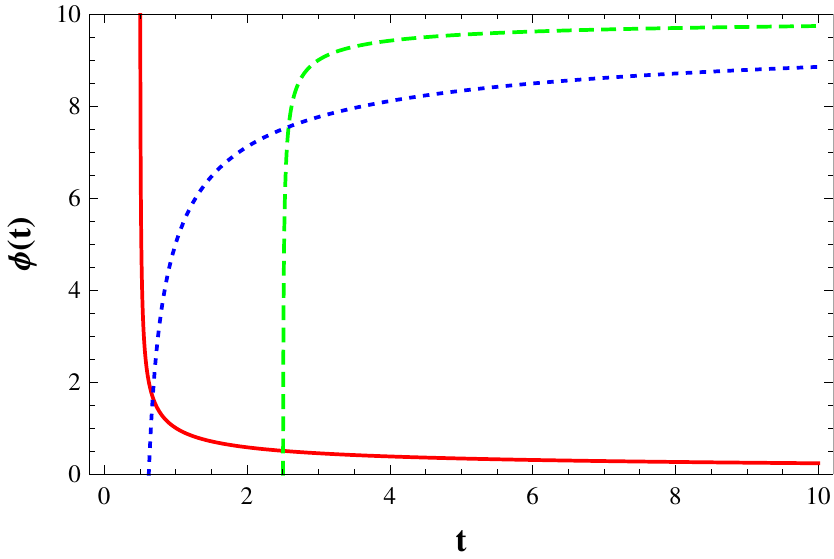}
\caption{Case I: time evolution of $\phi$. Dotted (blue) line stands for $\beta_1=1$, $\gamma_1=10$ and $\delta_1=5$, while for the dashed (green) line and solid (red) line, $\beta_1=5$, $\gamma_1=10$ and $\delta_1=1$ and $\beta_1=1$, $\gamma_1=0$ and $\delta_1=-1$, respectively.}
\label{fig1}
\end{figure}  

From (\ref{imm9}), the matter-energy density and pressure of the universe in this model read:

\begin{equation}\label{rho1}
\rho(t)=\frac{3\delta_1}{(\beta_1-2t)^{2}(\delta_1-\gamma_1\sqrt{2t-\beta_1})},
\end{equation}
\begin{equation}\label{p1}
p(t)=\frac{\delta_1}{(\beta_1-2t)^{2}(\delta_1-\gamma_1\sqrt{2t-\beta_1})}.
\end{equation}

\subsection{Case II: $a=a(t,l)$ and $\phi=\phi(t)$}\label{ss:ptl}

Let us generalize the previous case. If one makes $\phi(t)\rightarrow\phi(t,l)$ in Eq.(\ref{imm1}), the Einstein tensor stays the same. Therefore, We will make $a(t)\rightarrow a(t,l)$ in Eq.(\ref{imm1}), as

\begin{equation}\label{ptl1}
dS^{2}=dt^{2}-a(t,l)^{2}(dx^{2}+dy^{2}+dz^{2})-\phi(t)^{2}dl^{2},
\end{equation}
i.e., it is assumed the extra coordinate also influences the scale factor values.

The Einstein tensor, in this case, reads

\begin{equation}\label{ptl1}
G_0^{0}=3\left(\frac{\dot{a}}{a}\right)^{2}+3\left\{\frac{\dot{a}}{a}\frac{\dot{\phi}}{\phi}-\frac{1}{\phi^{2}}\left[\frac{a''}{a}+\left(\frac{a'}{a}\right)^{2}\right]\right\},
\end{equation}
\begin{equation}\label{ptl2}
G_1^{1}=2\frac{\ddot{a}}{a}+\left(\frac{\dot{a}}{a}\right)^{2}+\frac{\ddot{\phi}}{\phi}+2\frac{\dot{a}}{a}\frac{\dot{\phi}}{\phi}-\frac{1}{\phi^{2}}\left[2\frac{a''}{a}+\left(\frac{a'}{a}\right)^{2}\right],
\end{equation}
\begin{equation}\label{ptl3}
G_2^{2}=G_3^{3}=G_1^{1},
\end{equation}
\begin{equation}\label{ptl4}
G_4^{4}=3\left[\frac{\ddot{a}}{a}+\left(\frac{\dot{a}}{a}\right)^{2}\right]-\frac{3}{\phi^{2}}\left(\frac{a'}{a}\right)^{2},
\end{equation}
\begin{equation}\label{ptl5}
G_0^{4}=G_4^{0}=3\left(\frac{a'}{a}\frac{\dot{\phi}}{\phi}-\frac{\dot{a}'}{a}\right),
\end{equation}
with $'\equiv\partial/\partial l$.

The IMM application in Equations (\ref{ptl1}), (\ref{ptl2}), (\ref{ptl4}) and (\ref{ptl5}) yields

\begin{equation}\label{ptl6}
\left(\frac{\dot{a}}{a}\right)^{2}=-\frac{\dot{a}}{a}\frac{\dot{\phi}}{\phi}+\frac{1}{\phi^{2}}\left[\frac{a''}{a}+\left(\frac{a'}{a}\right)^{2}\right],
\end{equation}
\begin{equation}\label{ptl7}
2\frac{\ddot{a}}{a}=-\frac{\ddot{\phi}}{\phi}-\frac{\dot{a}}{a}\frac{\dot{\phi}}{\phi}+\frac{1}{\phi^{2}}\frac{a''}{a},
\end{equation}
\begin{equation}\label{ptl8}
\frac{\ddot{a}}{a}+\left(\frac{\dot{a}}{a}\right)^{2}-\frac{1}{\phi^{2}}\left(\frac{a'}{a}\right)^{2}=0,
\end{equation}
\begin{equation}\label{ptl9}
\frac{\dot{\phi}}{\phi}=\frac{\dot{a}'}{a'}.
\end{equation}

Solving (\ref{ptl6})-(\ref{ptl9}) for $a(t,l)$ yields

\begin{equation}\label{ptl10}
a(t,l)=\alpha_2[(2\kappa-1)t]^{\frac{1}{2}\frac{3\kappa-1}{2k-1}}e^{\kappa l},
\end{equation}
with $\alpha_2$ an arbitrary constant and $\kappa$ the constant of separation of variables.

From solution (\ref{ptl10}), we can write the Hubble and deceleration parameters respectively as

\begin{equation}\label{ptl11}
H(t)=\frac{1-3\kappa}{1-2\kappa}\frac{1}{2t},
\end{equation}
\begin{equation}\label{ptl12}
q(t)=\frac{\kappa-1}{3\kappa-1},
\end{equation}
with the former being related to the recession velocity of galaxies as the Hubble law $v=H(t)r$, with $H=\dot{a}/a$, and the latter being defined so that $q<0$ represents an accelerated expansion of the universe, with $q=-\ddot{a}a/\dot{a}^{2}$.

In the next section, some cosmological features of Cases I and II of the geometric model described by metric (\ref{i1}) will be presented.

\section{Cosmological features}\label{sec:cf}

Let us analyse the solutions obtained in the last section from a cosmological perspective. We will start recalling solutions (\ref{rho1}) and (\ref{p1}) of Case I. Remarkably, the relation between these equations is a factor of $3$. The relation between $p$ and $\rho$ is known as equation of state (EoS) $\omega$ and it is well known that $\omega=1/3$ for the radiation-dominated era of the universe \cite{ryden/2003}.

Therefore, purely from geometry (Eq.(\ref{imm1})), relativistic matter could be described. Specifically, this is a radiation-dominated universe model. In fact, such an argumentation is confirmed from Eq.(\ref{imm7}) since in standard model of cosmology, $a(t)\sim \sqrt{t}$ is exactly the time proportionality of the scale factor in a radiation-dominated universe \cite{ryden/2003}.

It is also possible to corroborate Eq.(\ref{imm9}) as related to a scalar field which describes the radiation-dominated universe. On this regard, \cite{ferreira/1998} have constructed a cosmological model from a primordial scalar field. The solution they have obtained for the scalar field was $\sim1/\sqrt{t}$, in agreement with (\ref{imm9}).

There is another form of indicating the correlation between solution (\ref{imm9}) and the radiation era, by analysing Fig.\ref{fig1}. By firstly checking the dotted (blue) and dashed (green) curves, one realizes that for small values of time, the field $\phi(t)$ still does not act. In fact, if a scalar field is related to the dynamics of a radiation-dominated universe, it is not expected to operate until the end of the inflationary era \cite{guth/1981}, i.e., the inflaton decaying. On the other hand, the solid (red) curve could be related to the ``graceful exit'' question \cite{modak/2012,kamada/2013}, which is the necessity of the inflaton to continuously decay into radiation. In this way, the high values of $\phi(t)$ for low values of $t$ would be a consequence of the initial values of $\phi$ being related to the decaying values of the inflaton. 

It is interesting and valuable to remark the following: although the electromagnetic potential was taken as null, an electromagnetic radiation solution was obtained in Case I. This implies that in KK theory, the extra coordinate might also retain some information on the electromagnetic force.

\begin{figure}[ht!]
\vspace{0.3cm}
\centering
\includegraphics[height=5cm,angle=00]{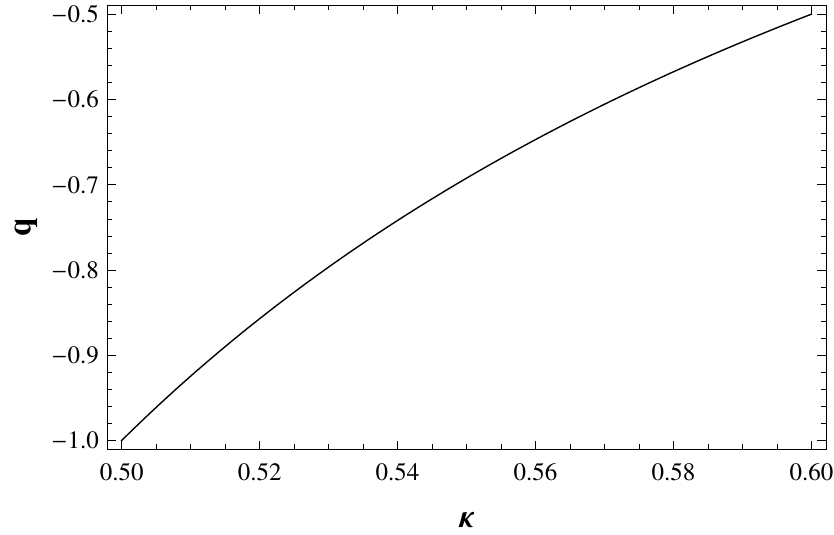}
\caption{Case II: $q$ as a function of $\kappa$.}
\label{fig2}
\end{figure}

Now we will recall Case II of Section \ref{sec:immhf}. The model solution for the deceleration parameter is presented in Eq.(\ref{ptl12}). The deceleration parameter observed present value was recently constrained from the 192 ESSENCE SNe Ia data to be $q_0=-0.788\pm0.182$ (\cite{lu/2008}). From (\ref{ptl12}), it is shown in Fig.\ref{fig2} below the range of values of $\kappa$ which is in accordance with observations.

In this way, purely from the metric (\ref{ptl1}) of Case II, in which the scale factor is assumed to be sensitive to both $t$ and $l$, it is possible to predict an accelerated expansion for the universe ($q<$0). The accelerated expansion of the universe was predicted by Type Ia supernova observations in the late 90's of the last century (\cite{riess/1998,perlmutter/1999}) and is an essential dynamical feature for cosmological models.

Fig.\ref{fig2} also reveals that the fiducial value $q_0=-0.788$ is obtained at $\kappa=0.537$.

Once the value of $\kappa$ is constrained, one can plot the scale factor in terms of $t$ and $l$ as Fig.\ref{fig3} below.

\begin{figure}[ht!]
\vspace{0.3cm}
\centering
\includegraphics[height=5cm,angle=00]{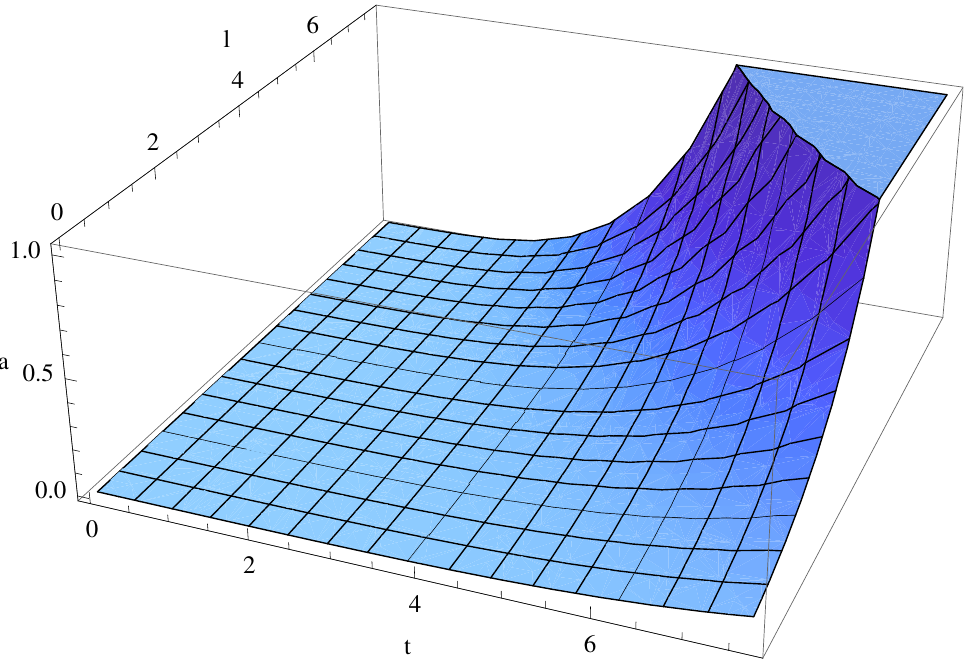}
\caption{Case II: $a$ as a function of $t$ and $l$ for $\alpha_2=1$ and $\kappa=0.537$.}
\label{fig3}
\end{figure}

Case II has also shown that the scalar field in the extra dimension may be the mechanism responsible not only for inducing matter in the observable universe, but also for inducing the counter-intuitive effect of accelerated expansion. That is, a scalar field ``living'' in the extra dimension together with a scale factor that ``feels'' the presence of such an extra dimension may be the responsible for the cosmic acceleration, evading the cosmological constant problem of standard cosmology \cite{weinberg/1989,padmanabhan/2003,hinshaw/2013}.

From Fig.\ref{fig3}, one can note that the value of $l$, i.e., the length scale of the extra coordinate, has a profound influence on the evolution of the scale factor. One realizes that the smaller the value of $l$, the greater the age of the universe, which is the value of $t$ for $a(t,l)=a_0=1$ (the present value of the scale factor \cite{ryden/2003}). According to observations of fluctuations on the temperature of cosmic microwave background radiation \cite{hinshaw/2013}, the present age of the universe is $\sim1.4\times10^{10}yr$. When inserting the natural constants in Case II, constrains to $l$ might be obtained. One expects such constrains to respect others, obtained by completely different approaches \cite{kolb/1986,arkani-hamed/1998,argyres/1998}.

\section{Discussion}

In the present article, we have constructed Kaluza-Klein models purely from the geometrical sector of a gravitational theory. The material content of the models have arisen as a geometrical manifestation of an extradimensional empty space-time.

Particularly, we have investigated the importance of scalar fields as extradimensional components of the metric. For the sake of generality, two cases for the metric (\ref{i1}), named $a=a(t)$ and $a=a(t,l)$. In both cases, $A_a=0$; models including non-null electromagnetic potentials might be reported as a forthcoming work.

Case I has shown to be able to describe relativistic matter as a geometrical manifestation of a 5D empty space-time. The features of a radiation-dominated universe could be observed in the scale factor solution, that evolves as a standard Friedmann-Robertson-Walker universe dominated by radiation. While in standard cosmology, the solutions for a radiation-dominated era are obtained from the assumption of an EoS like $p=\rho/3$, in the present model, such an EoS was obtained (rather than assumed), as one can check Eqs.(\ref{rho1})-(\ref{p1}). Beyond that, the solution (\ref{imm9}) for the scalar field responsible for inducing matter in the 4D observable universe agrees with that obtained in \cite{ferreira/1998}, for which the authors have constructed a cosmological model from a primordial scalar field. 

In Case II, the induced cosmological solutions led to an accelerated expansion of the universe. This was reflected in the negative values of $q$, which, for a range of values for the free parameter $\kappa$, was in agreement with observations \cite{lu/2008}. In this manner, this model has shown that it is possible to describe the cosmic acceleration as a purely geometrical effect, evading the cosmological constant problem (\cite{weinberg/1989,padmanabhan/2003,hinshaw/2013}). 

It is interesting to note that for $\kappa=0$ in Case II, the radiation-dominated universe is retrieved, since $q\rightarrow1$ as $\kappa\rightarrow0$ and $1$ is the deceleration parameter value at the radiation-dominated stage, according to standard cosmology (\cite{ryden/2003}).

The same happens if one applies the cylindrical condition in Case II. In fact, in KK theory, when recovering 4D gravity and Maxwell's equations from the cylindrical condition application to Eqs.(\ref{i1})-(\ref{i2}), the energy-momentum tensor induced by extradimensional geometrical terms is the one of electromagnetism, i.e., $T_{ab}^{EM}=g_{ab}F_{cd}F^{cd}/4-F_a^{c}F_{bc}$, with $F_{ab}\equiv\partial_a A_b-\partial_b A_a$. At the same time, it can be shown that the cylindrical condition application yields the EoS of radiation \cite{overduin/1997}. Such features were obtained in the present model, as one can see that Case I can be obtained from the cylindrical condition application in Case II.

\textcolor{black}{We would like to emphasize that by using scalar fields as extradimensional metric components we can open a new window to explain some standard cosmology issues. Therefore, in order to deal with these questions, we shall present in near future an approach in which the line element has a background with two interacting scalar fields. In this case, we hope that from such an inclusion, a greater number of new cosmological effects and phenomenological observations can be described.}

\acknowledgments

PHRSM would like to thank S\~ao Paulo Research Foundation (FAPESP), grant 2015/08476-0, for financial support. RACC is partially supported by FAPESP (Foundation for Support to Research of the State of S\~ao Paulo) under grants numbers 2016/03276-5 and 2017/26646-5.

\end{document}